\documentclass[lettersize,journal]{IEEEtran}
\usepackage{amsmath,amsfonts}
\usepackage{algorithmic}
\usepackage{algorithm}
\usepackage{array}
\usepackage[caption=false,font=normalsize,labelfont=sf,textfont=sf]{subfig}
\usepackage{textcomp}
\usepackage{stfloats}
\usepackage{url}
\usepackage{verbatim}
\usepackage{graphicx}
\usepackage{cite}
\usepackage{times}
\usepackage{comment}
\usepackage{tabularx}
\usepackage{epsfig}
\usepackage{pifont}
\usepackage{amssymb}
\usepackage{hyperref}
\usepackage{multirow}
\begin{document}

\title{Multi-Scale Feature Prediction with Auxiliary-Info\\for Neural Image Compression}

\author{Chajin Shin$^{*}$, Sangjin Lee$^{*}$, Sangyoun Lee,~\IEEEmembership{Member, IEEE,}
        % <-this % stops a space
\thanks{Manuscript received XX XX, XXXX; revised XX XX, XXXX}% <-this % stops a space
\thanks{$^{*}$Both authors contributed equally to this work.}}

\markboth{}%
{Shell \MakeLowercase{\textit{et al.}}: A Sample Article Using IEEEtran.cls for IEEE Journals}


\maketitle

\begin{abstract}
% Recently, image compression has shown significant improvements in rate-distortion performance with deep-learning techniques.
Recently, significant improvements in rate-distortion performance of image compression have been achieved with deep-learning techniques.
% A key factor contributing to this success is predicting an approximation of a latent vector, which is the output of the encoder, through another neural network using additional bits.
A key factor in this success is the use of additional bits to predict an approximation of the latent vector, which is the output of the encoder, through another neural network.
Then, only the difference between the prediction and the latent vector is coded into the bitstream, along with its estimated probability distribution.
We introduce a new predictive structure consisting of the auxiliary coarse network and the main network, inspired by neural video compression.
The auxiliary coarse network encodes the auxiliary information and predicts the approximation of the original image as multi-scale features.
The main network encodes the residual between the predicted feature from the auxiliary coarse network and the feature of the original image.
To further leverage our new structure, we propose Auxiliary info-guided Feature Prediction (AFP) module that uses global correlation to predict more accurate predicted features.
Moreover, we present Context Junction module that refines the auxiliary feature from AFP module and produces the residuals between the refined features and the original image features.
Finally, we introduce Auxiliary info-guided Parameter Estimation (APE) module, which predicts the approximation of the latent vector and estimates the probability distribution of these residuals.
We demonstrate the effectiveness of the proposed modules by various ablation studies.
Under extensive experiments, our model outperforms other neural image compression models and achieves a 19.49\% higher rate-distortion performance than VVC on Tecnick dataset.
\end{abstract}


\section{Introduction}
\IEEEPARstart{W}{ith} the increasing demand for high-resolution and high-quality images, there is a significant load on server storage and bandwidth for communications.
In response to this challenge, image compression is one of the most important tasks in image processing technology.
It drastically reduces file sizes while preserving quality, and extensive research has been conducted to achieve better rate-distortion performance.
Traditional lossy image compression methods include JPEG~\cite{JPEG}, JPEG2000~\cite{JPEG2000}, BPG~\cite{BPG}, and VVC intra~\cite{vvc}.
They divide the image into multiple blocks and utilize transformation, quantization, and entropy coding to eliminate redundant spatial information with low distortion.
However, because they are handcrafted methods, they are not fully optimized and cannot exploit complex non-linear operations.

Recently, deep learning-based image processing has emerged and shown remarkable performance improvements in various tasks~\cite{li2019filternet,pan2022real,liu2022deep,mao2022deep,li2023lightweight,wang2023asymmetric}.
The application of this technique to image compression has enabled significantly better rate-distortion performance compared to traditional image compression methods.
There are two main factors for this performance improvement.
The first is using non-linear transformation, which replaces traditional transformations, such as Discrete Cosine Transform (DCT). It converts the pixels of an image into a latent vector, effectively concentrating the information of the image.
Most neural image compression methods~\cite{end-to-end,end-to-end-attention,joint,asymmetric,shi2022variable,tang2022joint,li2022learned,channel-wise,hyperprior} are based on the structure of convolutional variational autoencoders (VAEs), as shown in Fig.~\ref{Fig:main_difference}-(a).
In this structure, the encoder performs the transformation and the decoder executes the inverse transformation.
The second factor is predicting an approximation of the latent vector using another neural network that utilizes additional side bits, subtracting the prediction from the latent vector to obtain the latent residual.
Then, this residual is encoded into the bitstream, along with its probability distribution estimated by a neural network.
Ball{\'e} et al.~\cite{hyperprior} propose a hyperprior that uses additional side information to model the probability distribution of the latent vector as a Gaussian distributions.
Minnen et al.~\cite{joint} not only estimate the probability distribution but also predict the approximation of the latent vector. Then, they subtract this prediction from the latent vector to store only the latent residual.
Moreover, they sequentially store the quantized latent vector, utilizing the already stored segments of the quantized latent vector to predict the subsequent segment to be stored. This approach leads to smaller residuals and a more accurate estimation of probability distributions.
Other works~\cite{lic-tcm,econtextformer,channel-wise,stf} introduce various structures to predict the approximation of the latent vector and probability distribution with a channel-wise auto-regressive manner or by using a transformer~\cite{swin-T}.

Neural Video Compression includes another prediction, namely, a temporal prediction.
In \cite{sheng2024spatial,dcvc,dcvc-tcm,dcvc-hem,dcvc-fm}, as shown in Fig.~\ref{Fig:main_difference}-(b), the motion vectors $m$ between reference frame $\hat{x}_{t-1}$ and the current frame $x_t$ are predicted and stored.
These motion vectors are used to warp the reference frame $\hat{x}_{t-1}$ to predict the current frame in the motion compensation module. Subsequently, the residuals are obtained implicitly by using a neural network that concatenates the features of the prediction frame and the current frame $x_t$.

Inspired by neural video compression structures, we introduce a new prediction architecture for neural image compression.
Specifically, we compress auxiliary information and predict the approximation of the original image as multi-scale features by the auxiliary coarse network, as illustrated in Fig.~\ref{Fig:main_difference}-(c).
These multi-scale features are concatenated with the features of the original image to implicitly obtain the feature residuals in the encoder of the main network.
In the decoder of the main network, the feature residuals are combined with the predicted multi-scale features to perform an inverse transform and get the reconstructed image.
To further exploit the new predictive structure, we propose Auxiliary info-guided Feature Prediction (AFP) module, which utilizes the global correlation of the auxiliary features to improve the prediction accuracy of the original image.
Furthermore, we present Context Junction module that comprises two sub-modules.
The first sub-module is Auxiliary-info Refiner, which combines the auxiliary feature with the main network feature and refines the auxiliary feature according to cross similarity with the combined feature.
The second sub-module, Auxiliary-info Subtractor, effectively and implicitly subtracts the refined feature from the original image, utilizing local as well as global correlation.
Finally, we propose Auxiliary info-guided Parameter Estimation (APE) module, which splits the latent vectors into multiple segments and sequentially predicts their approximation and the probability distribution with the auxiliary information.
Then, this module encodes each segment into the bitstream.
By conducting extensive experiments across various datasets, we demonstrate substantial rate-distortion performance improvement, where the proposed model outperforms VVC by 19.49\% on the Tecnick dataset.

The main contributions can be summarized as follows

\begin{itemize}
\item We utilize auxiliary information to predict the approximation of the original image as multi-scale features, and the main network implicitly subtracts the multi-scale features from the features of the original image to encode only the residual.
\item AFP module, which effectively predicts the original image using global correlation, is introduced.
\item We present Context Junction module, which refines predicted features and implicitly subtracts them from the features of the original image.
\item APE module is proposed to predict the latent vector and probability distribution with the auxiliary information.
\end{itemize}

\section{Related Works}
\label{related}
\subsection{Image Compression}
\noindent\textbf{Traditional Codec:} There exist various traditional image codecs to reduce network traffic and storage capacity loads, including JPEG~\cite{JPEG}, JPEG2000~\cite{JPEG2000}, BPG~\cite{BPG}, and VVC intra~\cite{vvc}.
To effectively reduce the spatial redundant information, the entire image is divided into multiple blocks of various sizes based on the contents of the image.
Subsequently, intra-prediction is performed to obtain the residuals.
Then, by using transformations such as DCT, these residuals are transformed into a domain where information can be effectively concentrated, followed by quantization.
Finally, entropy coding is employed to generate a bitstream.

\noindent\textbf{Learning-based:} Deep learning-based image processing methods have emerged and achieved significant performance improvement in many computer vision areas.
Recently, there are many attempts to apply these methods to image compression, achieving better rate-distortion performance than even the latest traditional codecs, such as VVC intra.
There are two important factors contributing to this dramatic increase in performance.

The first factor is utilizing the non-linear transformation that maps the pixels of an image into the latent vector $y$ that concentrates information of the image, replacing conventional transformations such as DCT.
Some works~\cite{end-to-end,easn} introduce non-linear adaptive activation or normalization.
They adaptively activate the intermediate features based on the contents, thereby transforming the input image into the latent vector more effectively.
Other works~\cite{gmm,end-to-end-attention} utilize an additional attention module to emphasize the important parts of the features and deactivate unnecessary parts.
However, these approaches only consider the correlation within local regions.
Recently, some studies~\cite{graph_attention,lic-tcm,stf} demonstrate that images have redundancy in local areas as well as globally.
To consider both local and global information, they utilize transformer structures and achieve significant performance improvements.

The second factor is predicting an approximation of the latent vector using another neural network with additional side bits.
This prediction is subtracted from the latent vector to obtain the latent residual.
Another neural network also estimates and models the probability distribution of the latent residual as the Gaussian or Laplace distribution.
Ball{\'e} et al.~\cite{hyperprior} introduce a hyperprior that utilizes additional side bits for estimating $\sigma$ to model the probability distribution of the latent vector $y$ as the Gaussian distribution, $\mathcal{N}(0, \sigma^2)$.
This approach enables the calculation of the bitrate of the latent vector, using it as a loss function, and facilitates adaptive entropy coding according to content, consequently showing significant performance improvement.
Minnen et al.~\cite{joint} predict $\mu$, which is an approximation of the latent vector, using additional bits.
They subtract $\mu$ from the latent vector, $y$, and perform quantization, $Q$, to obtain the latent residual, $\hat{r} = Q(y-\mu)$.
Thereafter, they apply entropy coding to the latent residual, assuming the Gaussian distribution with the estimated probability distribution.
In the decoder, the latent residual $\hat{r}$ are added to prediction $\mu$ to produce the quantized latent vector $\hat{y}$.
Moreover, they sequentially store the quantized latent vector across the spatial axis and utilize the already stored segments $\{..., \hat{y}_{i-2}, \hat{y}_{i-1}\}$ of the latent vector to predict the subsequent part $y_i$ to be stored.
This method significantly reduces spatial redundancy and increases performance by allowing smaller residuals and more accurate probability distributions.
Further, some works~\cite{econtextformer, entroformer} leverage a transformer to utilize the dependency of the spatio-channels axis or spatially long-term correlations to reduce the residual and estimate a more accurate probability distribution of the residual.

\subsection{Neural Video Compression}
Neural video compression utilizes reference frames to compress the current frame by removing both spatial and temporal redundancies.
Many works~\cite{dvc,elf-vc,lin2022dmvc,sheng2024spatial} predict the optical flow between the current frame and previously compressed frames used as references.
This flow is first compressed and then used to warp the reference frame to predict the current frame.
The difference between the current frame and predicted frame is compressed using an image compression method.
However, Li et al.~\cite{dcvc} demonstrate that a simple subtraction operation to remove redundancy between frames is not optimal.
Instead of a direct subtraction operation in the pixel domain, they concatenate it with the encoder's features across various scales, allowing the neural network to implicitly find a better optimized operation.
This approach achieves significantly higher performance improvements than simple subtraction operations.
Inspired by the structure of neural video compression, we propose a new prediction architecture for neural image compression. We first store auxiliary information in the auxiliary coarse network and then implicitly subtract it from the original image at the multi-scale feature levels to store only the feature residual in the main network.

\section{Method}
Our objective is to obtain latent vectors $z_\text{aux}$, $y_\text{aux}$, $z$, and $y$ that effectively concentrate the information of the original image $x$.
These latent vectors are then encoded into the bitstream by quantization and arithmetic coding.
We also aim to achieve a reconstructed image $\hat{x}$ with minimal distortion using the quantized latent vectors $\hat{z}_{\text{aux}}$, $\hat{y}_{\text{aux}}$, $\hat{z}$, and $\hat{y}$.
In this section, we first describe the overall structure, dividing it into an auxiliary coarse network and a main network. Then, we introduce the details of the proposed modules for each part.

\subsection{Overall Structure}
\subsubsection{Auxiliary Coarse Network}
The overall structure of the auxiliary coarse network is illustrated in Fig.~\ref{Fig:aux_coarse_network}.
The auxiliary coarse network takes the original image to compress the auxiliary information and predicts the approximation of the original image as multi-scale features.

The encoder comprises a convolutional layer with a kernel size of $4\times4$ and employs EASN~\cite{easn} for adaptive non-linear activation function.
At a 1/4 scale, we utilize Auxiliary info-guided Feature Prediction (AFP) module to predict the approximation of the original image features more accurately using global correlation.
This encoder transforms the original image into the latent vector, $y_\text{aux}$.
To concentrate the information of $y_\text{aux}$, an bitrate loss function, $R$, is used during training.
This loss function minimizes the bitrate of the quantized latent vector, $\hat{y}_\text{aux}$, by utilizing its estimated probability distribution $P$ as follows:

\begin{equation}
    R = -\mathbb{E}[log_2 P].
\end{equation}

To estimate the probability distribution of $\hat{y}_\text{aux}$, the side information, $\hat{z}_\text{aux}$, is utilized.
The hyper-encoder takes $y_\text{aux}$ to produce $z_\text{aux}$, and then quantization is applied to obtain $\hat{z}_\text{aux}$.

\begin{equation}
\begin{aligned}
    y_\text{aux} &= E_\text{aux}(x:\phi_{E_\text{aux}}) \\
    z_\text{aux} &= HE_\text{aux}(y_\text{aux}:\phi_{HE_\text{aux}}) \\
    \hat{z}_\text{aux} &= Q(z_\text{aux}),
\end{aligned}
\end{equation}

\noindent where $E_\text{aux}$ and $HE_\text{aux}$ represent the encoder and the hyper-encoder of the auxiliary coarse network, respectively. $Q$ denotes the quantization operation. $\phi_{E_\text{aux}}$ and $\phi_{HE_\text{aux}}$ are the optimized parameters of the encoder and the hyper-encoder, respectively.
$\hat{z}_\text{aux}$ is encoded into a bitstream using the lossless method of arithmetic coding.
Because $\hat{z}_\text{aux}$ does not utilize any priors, a factorized density model~\cite{hyperprior} $\psi_\text{aux}$ is used to estimate its probability distribution as follows:

\begin{equation}
    p_{\hat{z}_\text{aux}|\psi_\text{aux}}(\hat{z}_\text{aux}|\psi_\text{aux}) = \prod_{j}(p_{\hat{z}_{\text{aux}, j|\psi_\text{aux}}}(\psi_\text{aux}) * \mathcal{U}(-\frac{1}{2}, \frac{1}{2}))(\hat{z}_{\text{aux},j}),
\end{equation}

\noindent where $\hat{z}_{\text{aux},j}$ represents the $j$-th element of $\hat{z}_\text{aux}$. $\mathcal{U}$ and $*$ denote the uniform random distribution and convolution operation, respectively.
Thereafter, $\hat{z}_\text{aux}$ is fed into the hyper-decoder to obtain $z_\text{apm}$.

To estimate the probability distribution of the quantized latent vector, $\hat{y}_\text{aux}$, more accurately, we divide $y_\text{aux}$ into $2N_p$ segments.
For the $i$-th segment, we predict two key features: $\mu_{\text{aux},i}$, representing the approximation of the latent vector $y_{\text{aux},i}$, and $\sigma_{\text{aux},i}$, indicating the standard deviation of the Gaussian distribution.

\begin{equation}
    \mu_{\text{aux},i}, \sigma_{\text{aux},i} = PE(\hat{y}_{\text{aux}, <i}, z_\text{apm}),
\end{equation}

\noindent where $PE$ and $<i$ denote Parameter Estimator and $\{0, 1, ..., i-1\}$, respectively.
The approximation, $\mu_{\text{aux},i}$, is subtracted from $y_{\text{aux},i}$, followed by quantization to obtain the latent residual $\hat{r}_{\text{aux},i}$.
The quantized latent vector, $\hat{y}_{\text{aux}, i}$, is obtained by adding $\mu_{\text{aux},i}$ to the latent residual $\hat{r}_{\text{aux},i}$.

\begin{equation}
\begin{aligned}
    \hat{r}_{\text{aux},i} &= Q(y_{\text{aux},i} - \mu_{\text{aux},i}) \\
    \hat{y}_{\text{aux}, i} &= \hat{r}_{\text{aux},i} + \mu_{\text{aux},i}.
\end{aligned}
\end{equation}

\noindent Subsequently, we model the probability distribution of $\hat{r}_{\text{aux}}$ as the Gaussian distribution, characterized by a mean of $0$ and a standard deviation of $\sigma_{\text{aux},i}$.

\begin{equation}
    p_{\hat{r}_\text{aux}}(\hat{r}_\text{aux}|\hat{z}_\text{aux}) = \prod_j(\mathcal{N}(0,\sigma_{\text{aux},j}^2) * \mathcal{U}(-\frac{1}{2}, \frac{1}{2}))(\hat{r}_{\text{aux},j}).
\end{equation}

\noindent The details of the specific structure of the Parameter Estimator are described in Section~\ref{Subsection:APE_module}. Finally, the auxiliary coarse decoder uses $\hat{y}_{\text{aux}}$ to obtain the multi-scale prediction features ( $F_\text{pred}^{1\times}$, $F_\text{pred}^{1\times}$, $F_\text{pred}^{4\times}$, $F_\text{pred}^{16\times}$ ) corresponding to scales of 1/1, 1/2, 1/4, and 1/16, respectively.
The decoder has a structure symmetric to the encoder, consisting of a transposed convolutional layer with a kernel size of $4\times4$, EASN, and AFP module.

\subsubsection{Main Network}
The encoder of the main network subtracts the auxiliary information, obtained from the auxiliary coarse network, from the original image $x$ and transforms the feature residual into the latent vector $y$.
Conversely, the decoder combines the auxiliary information and the latent vector to reconstruct the original image.

In the encoder, the original image, $x$, is fed into a convolutional neural layer and then concatenated with the predicted feature $F_\text{pred}^{1\times}$. This is followed by EASN to implicitly obtain the feature residual.
This process is also executed at 1/2 scale using $F_\text{pred}^{2\times}$ with a convolutional layer of kernel size $4\times4$ and stride 2 for downsampling.
At the 1/4 scale, instead of using EASN, we use Context Junction module to effectively refine the predicted feature $F_\text{pred}^{4\times}$ and extract the feature residual.
Through these processes, the encoder produces the latent vector, $y$, which concentrates the information of the feature residual.

The side information, $\hat{z}$, is generated in the same manner as in the auxiliary coarse network, using $y$, hyper-encoder, quantization, and arithmetic coding.
Subsequently, $z_{pm}$ is obtained using the hyper-decoder and $\hat{z}$.
To effectively estimate the probability distribution of the latent vector $y$, we employ Auxiliary info-guided Parameter Estimation (APE) module, which has a structure similar to that of Parameter Estimator of the auxiliary coarse network. This module not only uses $z_{pm}$ but also incorporates the predicted feature $F_\text{pred}^{16\times}$ from the auxiliary coarse network.

\begin{equation}
    \mu_{i}, \sigma_{i} = APE(\hat{y}_{<i}, z_\text{pm}, F_\text{pred}^{16\times}).
\end{equation}

\noindent Similar to the auxiliary coarse network, we subtract the approximation, $\mu_i$, from the latent vector, $y_i$, and apply quantization to obtain the latent residual $\hat{r}_{i} = Q(y_{i} - \mu_{i})$.
Then, by adding $\mu_i$, we acquire a reconstruction $\hat{y}_{i} = \hat{r}_{i} + \mu_{i}$.
For the entropy loss function of $\hat{y}$, we model the probability distribution of $\hat{r}$ as the Gaussian with a mean of $0$ and a standard deviation of $\sigma$.

\begin{equation}
\label{eqn:main network probability distribution}
    p_{\hat{r}}(\hat{r}|\hat{z}, F_\text{pred}^{16\times}) = \prod_j(\mathcal{N}(0,\sigma_{j}^2) * \mathcal{U}(-\frac{1}{2}, \frac{1}{2}))(\hat{r}_{j}).
\end{equation}

\noindent Subsequently, the decoder of the main network takes the quantized latent vector, $\hat{y}$, and upsamples it using a transposed convolution with a $4\times4$ kernel size and EASN.
At the 1/4 scale, we exploit the Context junction module to refine the predicted feature $F_\text{pred}^{4\times}$ and combine it with the features from the main decoder.
At the 1/2 and 1/1 scales, we simply concatenate the predicted feature $F_\text{pred}^{2\times}$ and $F_\text{pred}^{1\times}$, and then feed them into EASN and a residual block, respectively.
Thereafter, we utilize a convolutional layer with a kernel size of $3\times3$ to obtain a reconstructed image $\hat{x}$.

\subsection{Evaluation}
We evaluate our models using five test datasets: the Kodak dataset~\cite{kodak}, which consists of images of size $768\times512$; the CLIC2021 Validation dataset~\cite{clic2021} and the CLIC2020~\cite{clic2020} Test (Professional and Mobile) dataset, which include images of various resolutions up to 2K; and the Tecnick~\cite{tecnick} dataset, with images of size $1200\times1200$.
To evaluate the rate-distortion performance, we use PSNR or MS-SSIM metrics to measure the distortion for each distortion function, $D$, and bits per pixel (bpp) to measure bitrates.

\subsubsection{Rate-Distortion Performance}
We compare the rate-distortion performance of our model with those of traditional codecs, including JPEG~\cite{JPEG}, JPEG2000~\cite{JPEG2000}, BPG~\cite{BPG}, and VTM~\cite{vvc}, which is VVC intra. Additionally, our comparisons extend to state-of-the-art (SoTA) neural image compression methods \cite{econtextformer, lic-tcm, stf, elic, easn, evc, entroformer, gmm, joint, hyperprior}.

\begin{figure*}[]
	\begin{center}
		\includegraphics[width=1\linewidth]{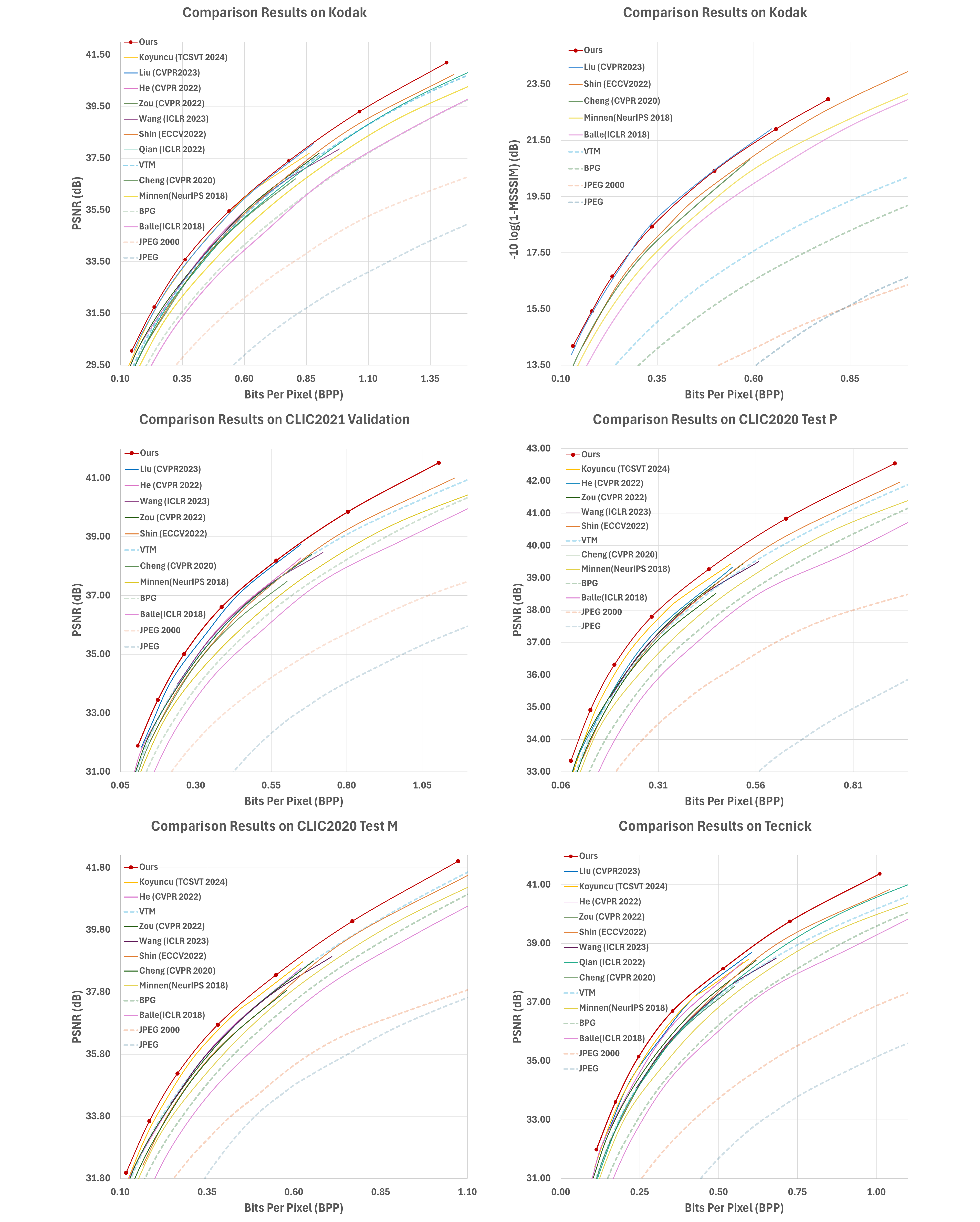}
		
	\end{center}
	\caption{Rate-Distortion Performance Comparison for both PSNR and MS-SSIM metrics.
	}
	\label{Fig:RD_performance}
\end{figure*}

Fig.~\ref{Fig:RD_performance} shows the rate-distortion performance plot, with bpp on x-axis and PSNR or MS-SSIM on the y-axis. The upper curve represents higher performance.
The comparison results on the Kodak dataset with PSNR distortion reveal that our model demonstrates superior performance over all other models across the entire bitrate range.
In case of MS-SSIM distortion loss function, we compare our model with other models that offer pretrained parameters or the exact values of each point.
Our model shows the same rate-distortion performance as the SoTA methods.
The Kodak dataset, which has a resolution of $768\times512$, does not closely represent the high resolutions of real-world images. Accordingly, considering the comparison results on CLIC2021 Validation, CLIC2020 Test (P for Professional, and M for Mobile), and Tecnick, which have a higher resolution, our model outperforms other models by a significant margin.

Table~\ref{tab:RD_rate} presents the rate-distortion performance using BD-rate~\cite{bdrate}, which indicates the percentage of bit reduction for the same distortion quality.
Thus, negative means bit saving. We calculate BD-rate using VTM as an anchor.
Our model shows much higher performance than other SoTA models with $13.79\%$ bit saving on the Kodak dataset.
In particular, for CLIC2020 Validation, CLIC2021 Test, or the Tecnick dataset, which are closer to real-world resolutions, our model achieves remarkable SoTA performance compared to any other method. The proposed model saves an average of $19.49\%$ bits for the same PSNR quality on the Tecnick dataset.

\begin{table*}[]
\caption{Rate-Distortion Performance with RD-rate.}
\label{tab:RD_rate}
\centering
\resizebox{\textwidth}{!}{
\begin{tabular}{c|c|ccccccccc}
\hline
Model                         & Dataset & VVC~\cite{vvc}  & Qian~\cite{entroformer}  & Shin~\cite{easn}  & Wang~\cite{evc}  & Zou~\cite{stf}   & He~\cite{elic}    
                              & Liu~\cite{lic-tcm}    & Koyuncu~\cite{econtextformer} & Ours   \\ \hline
                              & Kodak   & 0.00 & 1.65  & -0.28 & -0.71 & -3.30 & -4.89  & -11.49 & -11.81  & \textbf{-13.79} \\
\multirow{3}{*}{BD-rate (\%)} & CLIC2021 val    & 0.00 & OOM   & 1.43  & -1.29 & 0.09  & -4.41  & -12.21 & -       & \textbf{-15.97} \\
                              & CLIC2020 test-P   & 0.00 &  OOM & -0.31 & -1.89 & -3.66 & -6.18  & - & -11.96  & \textbf{-18.46} \\
                              & CLIC2020 test-M   & 0.00 &  OOM  & 4.15 & 0.64 & 0.25 & -1.37  & - & -7.03  & \textbf{-10.85} \\
                              & Tecnick & 0.00 & -0.45 & -2.96 & -1.70 & -6.21 & -10.91 & -14.36 & -13.90  & \textbf{-19.49}  \\ \hline
\end{tabular}
}
\end{table*}

\begin{figure}[]
	\begin{center}
		\includegraphics[width=0.99\linewidth]{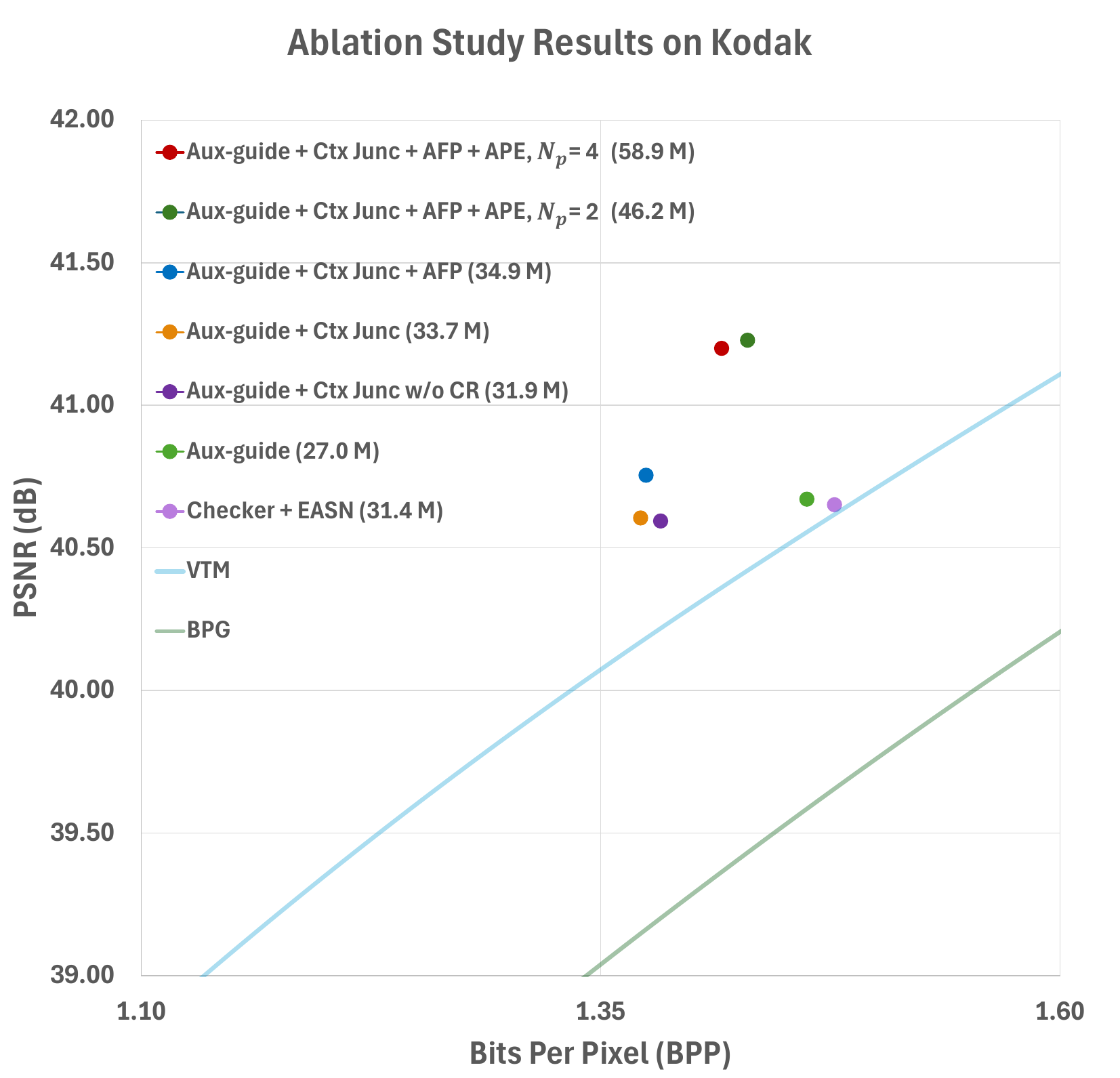}
		
	\end{center}
	\vspace{-0.5cm}
	\caption{Rate-Distortion Performance Comparison for Ablation Studies. The value in the parentheses indicates the total parameters of the models.
	}
	\label{Fig:ablation_results}
\end{figure}

\subsubsection{Qualitative Comparison}
\begin{figure*}[]
	\begin{center}
		\includegraphics[width=0.9\linewidth]{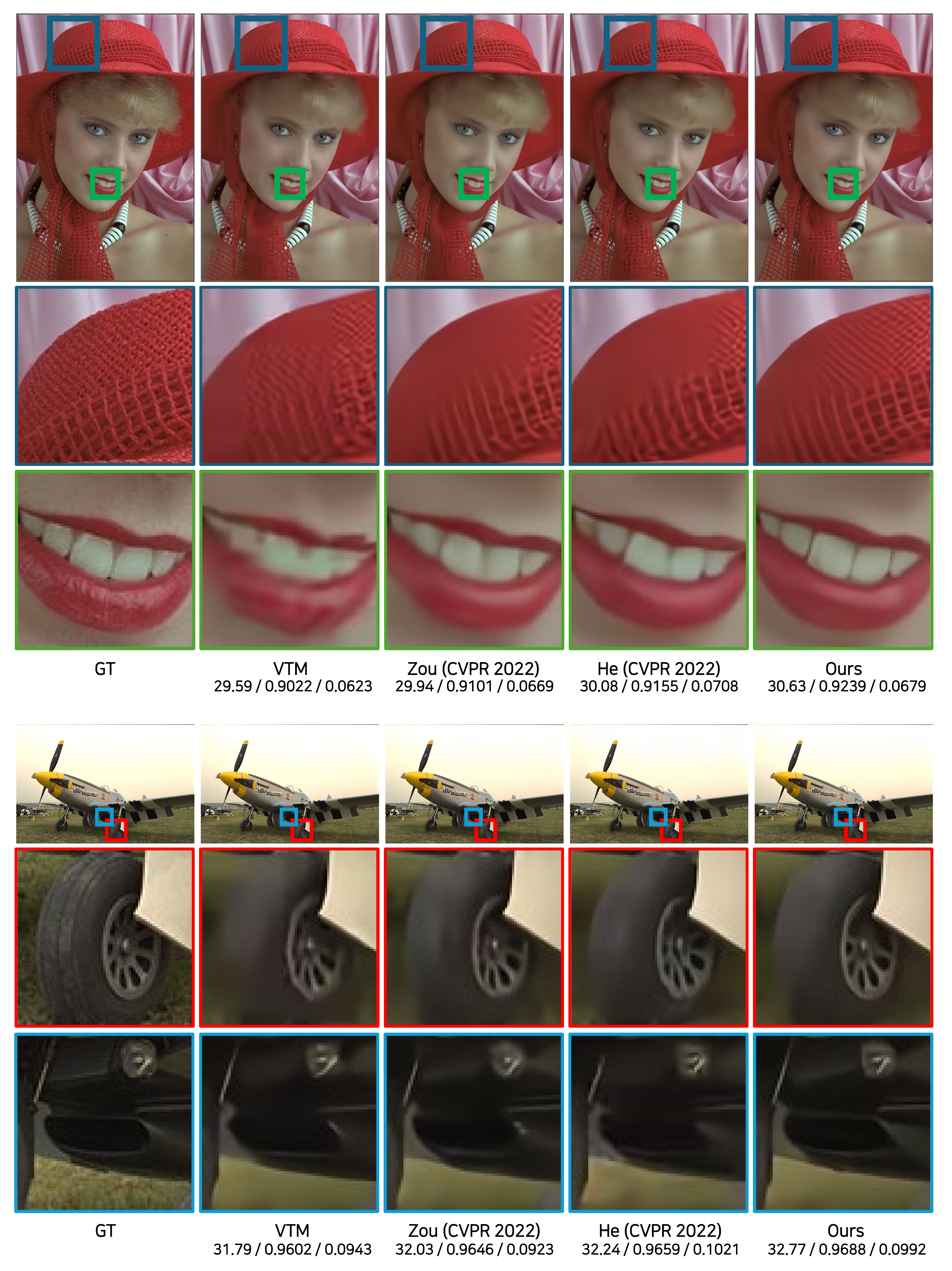}
		
	\end{center}
	\caption{Qualitative Comparison Results for \textit{Kodim04} (upper) and \textit{Kodim20} (lower) images from the Kodak dataset: Each value indicates PSNR / MS-SSIM / bpp, respectively.
	}
	\label{Fig:visualization}
\end{figure*}

We evaluate the visual quality of our model against the traditional VTM codec. Moreover, we compare with SoTA neural image compression methods~\cite{stf, elic} that provide pretrained parameters.
Fig.~\ref{Fig:visualization} presents the visual quality comparison results: \textit{Kodim04} is for the upper image and \textit{Kodim20} for the lower image, both from the Kodak dataset. Each value under the images presents PSNR / MS-SSIM / bpp, respectively.
As we can see in \textit{kodim04} image, the VTM shows block artifacts in the hat textures.
In addition, it distorts the structure in the teeth and lips.
Similarly, for neural image compression, both methods fail to capture the detailed texture of the hat and present incorrect structure in the teeth. By contrast, our model accurately depicts the detailed texture of the hat and maintains the correct structure of the teeth.
In the case of the \textit{Kodim20} image, our model successfully captures the precise structures of the wheel rim and holes, in contrast to other methods that struggle to preserve these details.
Furthermore, our model accurately reproduces the structure of the exhaust pipe, while other methods fails to capture the structure.
These qualitative comparison results indicate that our model outperforms other methods in preserving details and structure at similar bpps.

\subsection{Ablation Studies}
\subsubsection{Proposed Modules}
Fig.~\ref{Fig:ablation_results} represents the rate-distortion performance plot for ablation studies on the Kodak dataset. The value in parentheses represents the total parameters of the models.
The baseline, denoted as "Checker + EASN", comprises JA+EASN~\cite{easn} with the checkerboard context model~\cite{checkerboard}.
We modify the baseline to an Auxiliary Info-guided structure, referred to as "Aux-guide", achieving performance improvement with fewer parameters.
A dramatic increase in performance is observed upon integrating Context Junction module, even without Cross-info Refiner. Further performance increase is achieved by incorporating Cross-info Refiner into Context Junction module.
The addition of AFP module to the auxiliary coarse network also results in higher performance.
Our final model, which incorporates APE module with $N_p=4$, demonstrates the highest rate-distortion performance.
With these experimental results, we can confirm that the proposed modules effectively predict the entire image with auxiliary information and subtract them in the main network to store only the residual information.

\subsubsection{Auxiliary Information Ratio}
Table~\ref{tab:aux_byte_ratio} presents the ratio of auxiliary information bitstream bytes to the total bitstream byte size along with the compression ratio for the \textit{kodim04} image from the Kodak dataset.
At a high compression ratio, corresponding to a small $\lambda$, the auxiliary information occupies approximately $12\%$ of the total file size.
However, at a low compression ratio, which corresponds to a large $\lambda$, the ratio of auxiliary information increases dramatically.
This is because, at a high compression ratio, the reconstructed image primarily contains low-frequency components, which are easily predictable.
Thus, a small amount of the auxiliary information is sufficient to accurately predict the reconstructed image.
By contrast, at a low compression ratio, there are many high-frequency components.
The auxiliary information also carries some amount of high-frequency components to effectively predict the high-frequency components of the high-quality reconstructed image.

\begin{table*}[]
\caption{Ratio of Auxiliary Bytes to Total Bytes Size}
\label{tab:aux_byte_ratio}
\centering
\resizebox{0.8\textwidth}{!}{
\begin{tabular}{c|ccccccc}
\hline
$\lambda$         & 0.0025  & 0.005   & 0.01    & 0.02    & 0.04    & 0.08    & 0.16    \\ \hline
PSNR (dB)     & 31.0279 & 32.4634 & 34.1374 & 35.7999 & 37.5893 & 39.4199 & 41.3467 \\
BPP           & 0.0806  & 0.1362  & 0.2258  & 0.3578  & 0.5546  & 0.8107  & 1.1638  \\
Aux Bytes     & 569     & 853     & 1417    & 1965    & 4429    & 10425   & 22005   \\
Main Bytes    & 3393    & 5845    & 9685    & 15624   & 22832   & 29424   & 35200   \\
Aux Ratio (\%) & 14.36   & 12.73   & 12.76   & 11.17   & 16.25   & 26.16   & 38.47   \\ \hline
\end{tabular}
}
\end{table*}

\section{Conclusion}
In this paper, we introduce a new architecture for image compression that consists of the auxiliary coarse network and the main network. The auxiliary coarse network predicts the original image as multi-scale features, and the main network implicitly subtracts the prediction from the original image and encodes the residuals.
To further leverage this architecture, we propose Auxiliary info-guided Feature Prediction (AFP) module to predict the original image more effectively as multi-scale features.
In addition, we present Context Junction module, which refines the auxiliary feature and subtracts them from the original image feature using both local and global correlation.
Finally, we introduce Auxiliary info-guided Parameter Estimator (APE) module to predict an approximation of the latent residuals and estimate their probability distribution.
Extensive experiments across various datasets demonstrate that our model achieves SoTA performance.

% \section*{Acknowledgments}
% This should be a simple paragraph before the References to thank those individuals and institutions who have supported your work on this article.

% {\appendix[Proof of the Zonklar Equations]
% Use $\backslash${\tt{appendix}} if you have a single appendix:
% Do not use $\backslash${\tt{section}} anymore after $\backslash${\tt{appendix}}, only $\backslash${\tt{section*}}.
% If you have multiple appendixes use $\backslash${\tt{appendices}} then use $\backslash${\tt{section}} to start each appendix.
% You must declare a $\backslash${\tt{section}} before using any $\backslash${\tt{subsection}} or using $\backslash${\tt{label}} ($\backslash${\tt{appendices}} by itself
%  starts a section numbered zero.)}



\end{document}